\documentstyle[12pt,fleqn]{article}
\input psfig
\def\double{\baselineskip 24pt \lineskip 10pt}

\def\be{\begin{equation}}
\def\ee{\end{equation}}
\def\bea{\begin{eqnarray}}
\def\eea{\end{eqnarray}}
\def\re#1{{[\ref{#1}]}}

\def\la{\mathrel{\mathpalette\fun <}}
\def\ga{\mathrel{\mathpalette\fun >}}
\def\fun#1#2{\lower3.6pt\vbox{\baselineskip0pt\lineskip.9pt
        \ialign{$\mathsurround=0pt#1\hfill##\hfil$\crcr#2\crcr\sim\crcr}}}
\def\pag{{{\langle P \rangle }}}
\begin{document}
\begin{titlepage}
\vspace*{-62pt}
\begin{flushright}
FERMILAB--PUB--93/066-A\\
March 1993
\end{flushright}
\vspace{1.0in}
\begin{center}
{\Large \bf Axion Miniclusters and Bose Stars}\\
\vspace{0.6cm}
\normalsize
Edward W.\ Kolb$^{(1),(2)}$ and Igor I.\ Tkachev$^{(1),(3)}$

\vspace{36pt}

{\em $^{(1)}$NASA/Fermilab Astrophysics Center\\
Fermi National Accelerator Laboratory, Batavia, IL~~60510

\vspace{18pt}

$^{(2)}$Department of Astronomy and Astrophysics, Enrico Fermi Institute\\
The University of Chicago, Chicago, IL~~ 60637

\vspace{18pt}

$^{(3)}$Institute for Nuclear Research of the Academy of Sciences of
Russia, Moscow 117312, Russia}
\end{center}

\vspace*{18pt}

\baselineskip=24pt

\begin{quote}
\normalsize
\hspace*{2em}
Evolution of inhomogeneities in the axion field around the QCD epoch is studied
numerically, including for the first time important non-linear effects.  It is
found that perturbations on scales corresponding to causally disconnected
regions at $T \sim 1 \, {\rm GeV}$ can lead to very dense axion clumps, with
present density $\rho_a \ga 10^{-8}\,{\rm g \, cm^{-3}}$. This is high enough
for the collisional $2a \rightarrow 2a$ process to lead to Bose--Einstein
relaxation in the gravitationally bound clumps of axions, forming Bose stars.
\vspace*{12pt}

PACS number(s):  98.80.Cq, 14.80.Gt, 05.30.Jp, 98.70.--f

\noindent
\small email: rocky@fnas01.fnal.gov; tkachev@fnas13.fnal.gov

\end{quote}

\normalsize

\end{titlepage}

\newpage
\double

The invisible axion is one of the best motivated candidates for cosmic dark
matter, despite being subject to strong cosmological and astrophysical
constraints on its properties ($10^{10}\, {\rm GeV} \la f_a \la 10^{12}\, {\rm
GeV}$ for the axion decay constant; $10^{-5}\, {\rm eV} \la m_a \la 10^{-3}\,
{\rm eV}$ for the axion mass) \re{mt90}.  As dark matter, axions would play a
role in the evolution of primordial density fluctuations and formation of large
scale structure. In addition to its generic properties, axions also have unique
features as dark matter.  For instance, large amplitude density fluctuations
produced on scales of the horizon at the QCD epoch \re{mt86} lead to tiny
gravitationally bound ``miniclusters'' \re{hr88}. It was found that the density
in miniclusters exceeds by ten orders of magnitude the local dark matter
density in the Solar neighborhood \re{hr88}. This might have a number of
astrophysical consequences, as well as implications for laboratory axion
searches \re{ps83}.

In previous studies, spatial gradients of the axion field in the equations of
motion were neglected. This is a reasonable assumption for temperatures below
the QCD scale where the evolution of coherent axion oscillations can be treated
as pressureless, cold dust. However, we find that just at the crucial time when
the inverse mass of the axion is approximately the size of the horizon,
gradient terms become important, and a full field-theoretical approach is
needed.  Here we present the results of a numerical study of the evolution of
the inhomogeneous axion field around the QCD epoch. Though we only consider
spherically symmetric configurations, the importance of the combined effect of
the field gradients and the non-liner attractive self interaction should also
occur if we relax spherical symmetry. The resulting axion clumps are much
denser than previously thought, reaching the critical conditions for Bose star
formation \re{it91}.

The axion field $\theta(x)$ is created during the Peccei-Quinn symmetry
breaking phase transition at $T \sim f_a$, uncorrelated on scales larger than
the horizon at this time \re{ll90}.  For $T \la f_a$, the field becomes smooth
on scales up to the horizon, $H^{-1}(T)$, where $H$ is the expansion rate.
This continues until $T=T_1 \approx 1 \, {\rm GeV}$ when the axion mass
switches on, i.e., when $m_a(T_1) \approx 3H(T_1)$.  Coherent axion
oscillations then transform fluctuations in the initial amplitude into
fluctuations in the axion density.

Since the initial amplitude of coherent axion oscillations on the horizon scale
$H^{-1}(T_1)$ is uncorrelated, one expects typical positive density
fluctuations on this scale will satisfy $\rho_a \approx 2 \bar{\rho}_a$, where
$\bar{\rho}_a$ is mean cosmological density of axions \re{hr88}. At the
temperature of equal matter and radiation energy density, $T_e = 5.5 \,\Omega_a
h^2 \, {\rm eV}$ \re{kt}, these fluctuations are already non-linear and will
separate out as miniclusters with  $\rho_a \approx 3\, (10\, {\rm eV})^4
\approx 10^{-14}\, {\rm g \, cm^{-3}}$  \re{hr88}.  The minicluster mass will
be of the order of the dark-matter mass within the Hubble length at temperature
$T_1$, $M_{\rm mc} \sim 10^{-9}\, M_\odot$.  The radius of the cluster is
$R_{\rm mc} \sim 10^{13}$cm, and the gravitational binding energy will result
in an escape velocity $v_e/c \sim 10^{-8}$. Note that the mean phase-space
density of axions in such a gravitational well is enormous: $n \sim \rho_a
m_a^{-4}v_e^{-3} \sim 10^{48}f_{12
}^4$, where $f_{12} \equiv f_a /10^{12}\, {\rm GeV}$.

We will show below that due to non-linear effects, a substantial number of
regions can have axion density at $T>T_e$ many times larger than $2\,
\bar{\rho}_a$.

Let us parametrize the energy density of a single fluctuation as
$\rho_a(T_e<T<T_1,\theta_i) \equiv 3\, \Phi(\theta_i) T_e  s/ 4$, where
$\theta_i$ is the misalignment angle at $T_1$, $s$ is the entropy density, and
$\Phi(\theta_i)=1$ corresponds to the mean axion density. The energy density
inside a given fluctuation  is equal to the radiation energy density at $T =
\Phi(\theta_i) T_e$. At that time the fluctuation becomes gravitationally
non-linear and collapses. Consequently, at $T_e$
\begin{equation}
\rho_a(\theta_i ) \sim \Phi^4(\theta_i) \bar{\rho}_a(T_e),
\label{rhofl}
\end{equation}
will be the minicluster density after it separates out as a bound object. Even
a relatively small increase in $\Phi(\theta_i)$ is important because the
density depends upon the fourth power of $\Phi(\theta_i)$.

Ref.\ \re{mt86} demonstrated that due to anharmonic effects for fluctuations
with $\theta_i$ close to $\pi$, some correlated regions can have  values of
$\Phi(\theta_i)$ larger than just a factor of two. The reason is simple: the
closer $\theta_i$ is to the top of the axion potential,
\begin{equation}
V(\theta ) = m_a^2(T)f_a^2 (1- \cos \theta ) \equiv \Lambda_a^{4}(T) (1- \cos
\theta ),
\label{pot}
\end{equation}
the later axion oscillations commence. However this effect alone is not very
significant. In the range $0.1 \la \xi \la 10^{-3}$ we can parametrize it as
$\Phi(\theta_i) \approx 1.5 (\theta_i /\pi)^2 \xi^{-0.35}$, where $\xi \equiv
(\pi-\theta_i)/\pi$, and $\Phi(\theta_i)$ is significantly larger than $2$ only
for field values very finely tuned to the top of the potential. Moreover, the
axion field is not exactly coherent on the horizon scale, and small
fluctuations might spoil this picture.

At temperatures $T \gg T_1$, the potential is negligible in the equations of
motion compared to the gradient terms which force the field to be homogeneous
on scales less than the horizon. At $T \ll T_1$, on the contrary, gradients can
be neglected and one can treat the evolution of fluctuations as that of a
pressureless gas. Clearly at $T \sim T_1$, both the gradient terms and the
potential are important, and in order to find the energy density profile at
freeze out one has to trace the inhomogeneous field evolution through the epoch
$T\sim T_1$.

It is convenient to work in conformal coordinates with metric $ds^2 =
a^2(\eta)(d\eta^2- d\vec{x}\,^2)$. During radiation dominance $a \propto \eta$
and $\eta \propto T^{-1}$. The dependence of the axion mass upon the
temperature at $T > \Lambda_{\rm QCD}$ can be found in the dilute-instanton-gas
approximation \re{gpy81}, and can be parametrized as a power law, $m_a^2(\eta)
= m_a^2(\eta_*) (\eta/\eta_*)^n$, where $n=7.4 \pm 0.2$ \re{mt86}. Introducing
the field $\psi=\eta\theta$, the equations of motion for a spherically
symmetric axion fluctuation in an expanding Universe is of the form
$\ddot{\psi} - \psi^{\prime \prime} -2\psi^\prime/r +\bar{\eta}^{n+3} \sin
\left( \psi/\eta \right) =0$, where $\bar{\eta}$ is the reduced conformal time
parameter $\bar{\eta}=\eta/\eta_*$, and $m_a(\eta_*)=H(\eta_*)$. The radial
coordinate $r$ is defined in the comoving reference frame, with $r=1$
corresponding to $R_{\rm phys}(\eta_*) = H^{-1} (\eta_*)$.

\begin{figure}[t]
\vspace*{12cm}
FIG.\ 1. Energy density contrast in a fluctuation with initial radius $r_0
=1.8$ and $\theta^<_i = 2.75$ at several moments of time as a function of
comoving radius $r$. The density contrast is normalized to the value of the
homogeneous energy density at $\bar{\eta}=4$.
\end{figure}

We integrated this equation numerically for a wide range of initial conditions.
 We evolved configurations which are at rest at $\bar{\eta}=0.1$.  The initial
distribution of the field can be parametrized by the initial radius of the
fluctuation, $r_0$, the initial value of the field inside, $\theta^<_i$, the
initial value of the field outside, $\theta^>_i$, and the width of transient
region, $\Delta r$. The important common feature is that the final density
distribution develops a sharp peak in the center. The larger the gradients of
initial configuration, the higher the final peak, e.g., the peak grows with
increase in $|\theta^<_i - \theta^>_i|$.  The peak also grows with decreasing
width of the transient region. We present here the results of runs with initial
amplitude of the field outside the fluctuation equal to the r.m.s.\ value of
the misalignment angle, i.e., $\theta^>_ i= \pi /\sqrt{3}$, and width of
transient layer  $\Delta r \sim 0.6$.

\begin{figure}[t]
\vspace*{12cm}
FIG.\ 2.  Energy density profiles at $\bar{\eta}=4$ for identical initial
fluctuations evolved with different Lagrangians. Solid line: axion case; dashed
line: $V(\theta )\propto \theta^2/2$; dotted line: $V(\theta )\propto
\theta^2/2 + \theta^4/4$; dash-dotted line: axion potential with field
gradients switched off.
\end{figure}

Energy density profiles as a function of time are presented in Fig.\ 1 for a
typical case.  At $\bar{\eta}=1$ there are two waves, incoming and outgoing,
both propagating with the velocity of light. At approximately $\bar{\eta}=2$
the incoming wave reaches the center and the outgoing wave reaches $r \approx
3.5$.  At later times the wave front does not move significantly because the
axion mass effectively switches on at $\bar{\eta} \approx 2$, and the edge of
the fluctuation ``freezes.''

One reason for energy density growth at later times is the continuing increase
of the axion mass. However the relative density contrast in the center with
respect to the unperturbed homogeneous environment continues to increase up to
the final time of integration, $\bar{\eta}=4$. This is entirely a non-linear
effect. One can see this in the following way: The average pressure over a
period of homogeneous axion oscillations in potential Eq.\ (\ref{pot}) is
negative, and is equal to $\pag\simeq-\Lambda_a^4(T) \theta_0^4 / 64$, where
$\theta_0$ is the amplitude of the oscillations \re{it86}. In other words, the
axion self-interaction is attractive. The larger the amplitude of oscillations
inside the fluctuation, the more negative will be the pressure inside, and
consequently, fluctuations with excess axions will contract in the comoving
volume. In addition, matter with a smaller pressure suffers less redshift in
the energy density. To see this effect we present in Fig.\ 2 the final density
profiles correspondin
g to identical initial field distributions evolved with different potentials:
the axion potential of Eq.\ (\ref{pot}); the axion potential with gradients
artificially switched off; a pure harmonic potential, $V(\theta )\propto
\theta^2/2$, where $\pag=0$; and the potential $V(\theta) \propto \theta^2/2 +
\theta^4/4$, where $\pag>0$.  Note that for the harmonic potential, at
$\bar{\eta}=4$ the maximal density excess is only about $3$, i.e., {\em ten}
times smaller than for the axion potential.

The dependence of the energy density contrast in the center upon the initial
radius is shown on Fig.\ 3. In the whole range of values of $r_0$ plotted, the
energy density takes its maximum value just in the center of the
 final configuration. Only if $r_0 < 1.55$ or $r_0 > 2.05$ does the final
energy density profile have a maximum at some non-zero radius. In a sense, the
initial radius of the fluctuation in the plotted range is more or less tuned in
such a way that the arrival of the incoming wave at the center is synchronized
with the switching on of the axion mass. However, there is nothing unnatural in
this ``synchronization,'' since as larger and larger scales enter the horizon
in an expanding Universe there will always be a scale for which the incident
wave of a disappearing fluctuation reaches the center just at the moment of
freeze out.

\begin{figure}[t]
\vspace*{12cm}
FIG.\ 3. Dependence of density contrast in the center of a fluctuation at
$\bar{\eta}=4$ upon the initial radius of a fluctuation for several values of
the initial misalignment angle inside the fluctuation.
\end{figure}

Quantitatively, the assumption of spherical symmetry is very important.
However, in general any isolated contrast in the initial misalignment angle
will decay via incoming and outgoing waves which will not possess spherical
symmetry. The overall picture will be the same as in the spherical case, but
the values of the maximal energy contrast in the final configuration at a given
$\theta_i^<$ will be smaller. Note in this respect, that the final density
contrast rapidly grows with increase of $\theta_i^<$ (see Fig.\ 3) due to the
attractive self-interaction resulting in negative pressure.  This has nothing
to do with the symmetry of the fluctuation, and we may expect to find large
density contrasts in regions where the field values happen to be close to $\pi$
initially \re{dw}.

The effect of the field gradients is important not only in the discussion of
the formation of high density peaks, but also in the careful estimate of the
mean density of axion matter. We found that the total excess mass of axions
within a fluctuation, compared to the  homogeneous background, does not vary
much, and is equal approximately to half of the excess mass if gradients in the
equations of motion would be neglected. This deficit might be attributed to the
redshift at early times, $\bar{\eta}\la 2$, when axions are still relativistic.

The energy density contrast plotted in Figs.\ 1 and 2 will coincide with the
factor $\Phi(\theta_i)$ in Eq.\ (\ref{rhofl}) if we assume that the mean
cosmological density of axions corresponds to homogeneous oscillations with
initial amplitude equal to the r.m.s.\ value of the misalignment angle. As we
have noted already, the energy density in an axion clump after it separates out
from the general expansion will be $\Phi^4(\theta_i)$ times larger than the
energy density at $T_e$.  So a density contrast of 30 will correspond to
roughly a factor of $10^6$ in the energy density of the cluster at $T<T_e$.

All axion miniclusters could be, in principle, relevant to laboratory axion
search experiments, since for a minicluster with $\Phi$ as small as 2, the
density is $10^{10}$ larger than the local galactic halo density. However, the
probability of a direct encounter with a clump is small. The interesting
question arises, could there be any astrophysical consequences of very dense
axion clumps? Below we shall discuss the possibility of ``Bose star'' formation
inside axion miniclusters.

The physical radius of an axion clump at $T_e$ is larger by many orders of
magnitude than the de Broglie wavelength of an axion in the corresponding
gravitational well. Consequently, gravitational collapse of the axion clump and
subsequent virialization can be described in the usual terms of cold dark
matter particles.  In a few crossing times some equilibrium (presumably close
to an isothermal) distribution of axions in phase space will be established. It
is remarkable that in spite of the apparent smallness of axion quartic
self-couplings,  $|\lambda_a| = (f_\pi/f_a)^4 \sim 10^{-53} f_{12}^{-4}$, the
subsequent relaxation in an axion minicluster due to $2a \rightarrow 2a$
scattering can be significant  as a consequence of the huge mean phase-space
density of axions \re{it91}. In the case of Bose-Einstein statistics the
inverse relaxation time is $(1+\bar{n})$ times the classical expression, or
$\tau_R^{-1} \sim  \bar{n}\, v_e \sigma \rho_a /m_a $, where $\sigma$ is the
corresponding cross section. For part
icles bounded in a gravitational well, it is convenient to rewrite this
expression in the form \re{it91}
\begin{equation}
\tau_R  \sim  m_a^7 \lambda_a^{-2}\rho_a ^{-2}v_e^2 .
\label{rl}
\end{equation}
The shallower the gravitational well at a given density of axions, the larger
the mean phase space density, and consequently the smaller the relaxation time
due to the $v_e^2$ dependence in Eq.\ (\ref{rl}). Note also the dependence of
the inverse relaxation time upon the square of the particle density.

The relaxation time (\ref{rl}) is smaller then the present age of the Universe
if the energy density in the minicluster satisfies
\begin{equation}
\rho_{10}  >  10^{6} v_{-8}\sqrt{f_{12}},
\label{rt}
\end{equation}
where  $\rho_{10} \equiv \rho /(10 \, {\rm eV})^4$ and $v_{-8} \equiv v_e
/10^{-8}$. If this occurs, then an even denser core in the center of the axion
cloud should start to form. An analogous process is the so-called gravithermal
instability caused by gravitational scattering.  This was studied in detail for
star clusters, where the ``particles" obey classical Maxwell--Boltzmann
statistics. Axions will obey Bose--Einstein statistics, with equilibrium
phase-space density $n(p)  = n_{\rm cond} + [ e^{\beta E} -1 ]^{-1}$,
containing a sum of two contributions, a Bose condensate and a thermal
distribution. The maximal energy density that non-condensed axions can saturate
is $\rho_{\rm ther} \sim m_a^4 v_e^3 $, which corresponds to $\bar{n}_{\rm
ther} \sim 1$. Consequently, given the initial condition $\bar{n} \gg 1$, one
expects that eventually the number of particles in the condensate will be
comparable to the total number of particles in the region where relaxation is
efficient. Under the influence of self-g
ravity, a Bose star [\ref{it86},\ref{rb69}] then  forms \re{it91}. One can
consider a Bose star as coherent axion field in a gravitational well, generally
with non-zero angular momentum in an excited energy state \re{it86}.

Comparing Eqs.\ (\ref{rhofl}) and (\ref{rt}), we conclude that the relaxation
time is smaller than the present age of the Universe and conditions for Bose
star formation can be reached in miniclusters with density contrast $\Phi
(\theta_i) \ga 30$ at the QCD epoch. For examples of such density contrasts,
see  Figs.\ 1 and 3.

Under appropriate conditions stimulated decays of axions to two photons in a
dense axion Bose star are possible [\ref{it86},\ref{it87}] (see also
\re{kw90}), which can lead to the formation of unique radio sources---axionic
masers. In view of results of present paper we conclude that the questions of
axion Bose star formation, structure and possible astrophysical signatures
deserves detailed study.

In conclusion, we have presented a numerical study of the evolution of
inhomogeneties in the axion field around the QCD epoch, including for the first
time important non-linear effects.  We found that the non-linear effects can
lead to a much larger core density of axions in miniclusters than previously
estimated. The increase in the density may be sufficiently large that axion
miniclusters might exceed the critical density necessary for them to relax to
form Bose stars.


It is a pleasure to thank  H.\ Feldman, J.\ Frieman, A.\ Kashlinsky,  A.\
Klypin,  D.\ Pogosyan,  A.\ Stebbins,  M.\ Turner, and R.\ Watkins for useful
discussions.  This work was supported in part by the DOE and NASA grant
NAGW-2381 at Fermilab.

\newpage

\centerline{\bf REFERENCES}

\frenchspacing
\begin{enumerate}
\item \label{mt90}
For recent reviews, see M. S. Turner, Phys. Rep. {\bf C197}, 67 (1990);
G. G. Raffelt, Phys. Rep. {\bf C198}, 1 (1990).
\item \label{mt86}
M. S. Turner, Phys. Rev. D {\bf 33}, 889 (1986).
\item \label{hr88}
C. J. Hogan and M. J. Rees, Phys. Lett. {\bf B205}, 228 (1988).
\item \label{ps83}
P. Sikivie, Phys. Rev. Lett. {\bf 51}, 1415 (1983).
\item \label{it91}
I. I. Tkachev, Phys. Lett. {\bf B261}, 289 (1991).
\item \label{ll90}
This does not necessarily require the reheating temperature after inflation to
be higher than $f_a$, since inflation itself can produce strong fluctuations in
the axion field as discussed in
A. D. Linde and D. H. Lyth, Phys. Lett. {\bf B246}, 353 (1990);
D. H. Lyth and E. D. Stewart, Phys. Rev. D {\bf 46}, 532 (1992).
\item \label{kt}
E. W. Kolb and M. S. Turner, {\em The Early Universe}, (Addison-Wesley, Redwood
City, Ca., 1990).
\item \label{gpy81}
D. Gross, R. Pisarski, and L. Yaffe, Rev. Mod. Phys. {\bf 53}, 43 (1981).
\item \label{it86}
I. I. Tkachev, Sov. Astron. Lett. {\bf 12}, 305 (1986).
\item \label{dw}
We will discuss in a future paper the possibility that collapsing domain walls
leave behind very dense clumps of non-relativistic axions.
\item \label{rb69}
R. Ruffini and S. Bonozzola, Phys. Rev. {\bf 187}, 1767 (1969);
J. D. Breit, S. Gupta, and A. Zaks, Phys. Lett. {\bf B140}, 329 (1984).
\item \label{it87}
I. I. Tkachev, Phys. Lett. {\bf B191}, 41 (1987).
\item \label{kw90}
T. W. Kephart and T. J. Weiler, preprint VAND-TH-90-2, 1990 (unpublished).

\end{enumerate}

\end{document}